\documentclass[aps,twocolumn,amssymb,showpacs,amsmath,preprintnumbers]{revtex4}
\usepackage{graphicx}
\usepackage{bm}

\begin{document}

\title{Elastic properties of vanadium pentoxide aggregates and topological
defects} 
\author{L. V.~Elnikova}
\affiliation{A. I. Alikhanov Institute for Theoretical and Experimental Physics, \\
25, B. Cheremushkinskaya st., 117218 Moscow, Russia}
\date{\today}

\begin{abstract}
We study the aqueous solution of vanadium pentoxide by using
topology methods. The experiments by Zocher, Kaznacheev, and Dogic
exhibited, that in the sol phases of $V_2O_5-H_2O$, the tactoid
droplets of $V_2O_5$ can coalesce. In the magnetic field, this
effect is associated with a gauge field action, viz. we consider
coalescence (in the topologically more convenient term, "junction")
of droplets as annihilation of topological defects, concerning with
the tactoid geometry. We have shown, that in the magnetic field, the
tactoid junction is mainly caused by non-Abelian monopoles
(vortons), whereas the Abelian defects almost do not annihilate.
Taking into account this annihilation mechanism, the estimations of
time-aging of the $V_2O_5-H_2O$ sols may be specified.
\end{abstract}
\maketitle
\section{Introduction}
The tactoid sol phase of the $V_2O_5-H_2O$ system has been
discovered at the 20-th years of the last century by Zocher (see
references in [1, 2]). At the beginning of our century, the tactoid
drops (tactoids) have been investigated on the optical experiments
by Kaznacheev [2], Lavrentovich [3], Dogic (see [4] and references
in [5]), and their coworkers. The tactoid phase is chemically
classified as the lyotropic inorganic nematic [1]. The tactoids
coexist with the isotropic liquid phase at the mass concentration of
$V_2O_5$, amounting 0.3-2.1 percents, and under other standard
conditions [2].

The thermodynamic parameters and $pH$cause the dynamics of their
formation, in particular, the junction.

The tactoid geometry is evolved complicatedly (and mutually
inversely) in depending on time-aging of the sols [2].

Due to the de Gennes's theory [6], the tactoid shape stabilization
is defined by competition between the elastic energy of the nematic
phase, the surface energy, and the anchoring energy [2]. The minimum
of the tactoid free energy provides an equilibrium shape of a
droplet. The measured macroscopic elastic moduli are in a very large
ratio ($\frac{K_3}{K_1}
> 100$), that distinguishes $V_2O_5-H_2O$ from other lyotropic
liquid crystals (LC), whose typical values of $\frac{K_3}{K_1}$ are
in order of ten.

In the magnetic field, the prolate droplets are aligned by their
long axes parallel to the field. Then the special case of the
junction of tactoid poles may be observed [1, 2].

Remarkably, that the sol phases of $V_2O_5-H_2O$ were conditionally
sorted on a shape polarity and a nematic director field [5] as of a
homogeneous and a non-uniform field, and of the spherical and the
bispherical [2] drops with boojums. Strikingly simultaneously, these
phases have been parsed (see [5, 7, 8] and references therein)
basing on the experiments by Dogic (references in [5]), performed
independently of Kaznacheev.

In this paper, we study the mesomorphism of the $V_2O_5-H_2O$ system
during the tactoid junction and specify the character of the
mesomorphic consequence there. Our goal is to define the influence
of junction onto dynamic parameters of the sol system, including
time-aging of the sols. In addition, aging of these sols in water is
an applied problem of ecology, since $V_2O_5$ contains in coal
impurities, generated in result of work of thermal power stations.

From a topological standpoint, poles of a tactoid are the point
defects, boojums. As will readily be observed, we have to do with a
quantum phase transition, the analogous topological singularities of
two poles (each admitting a flux) were announced by Haldane [9] for
the quantum Hall semiconductors. Also, there is a convenient analogy
with the boojum formalism for the superfluid phases of $^3He$ and
$^4He$ [10], however their varied topology descriptions does not
allow to explain the case of the tactoid coalescence.

\section{Formalism}

Geometry of the droplets obeys the local nematic order parameter $\textbf{n}$,
which is oriented relatively to a droplet surface (Fig. 1).

\begin{figure}
\includegraphics*[width=70mm]{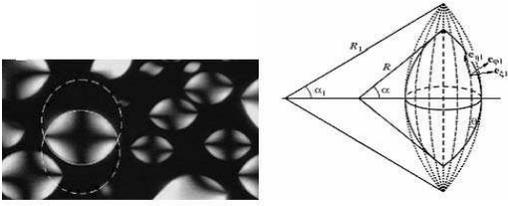}
\caption{\small The director field on the tactoid surface, taken
over [12]. $R_i$ and $\alpha$ are the geometric parameters,
$\gamma=(\frac{\tan(\alpha_1/2}{\tan(\alpha/2)})^2$,
$0\leq\gamma\leq1$, the vectors $\textbf{e}_i$
$i=\varphi_{kazn},\xi,\eta_{kazn}$ denote the bispherical
coordinates.}
\end{figure}
The free energy functional of a tactoid in the magnetic field is
summed up from the Frank elastic energy $F_{el}$ and the magnetic
energy $F_m$ [2, 5]:
\begin{equation}\label{freeen}
F=F_{el}+F_m,
\end{equation}
\begin{equation}\label{long}
F_{el}=\int_{V}d^{3}{\mathbf r}[\frac{K_1}{2}( \nabla \cdot{\mathbf
n})^{2} + \frac{K_2}{2} ({\mathbf
n}\cdot \nabla \times {\mathbf n})^{2} + \\
\frac{K_3}{2}[{\mathbf n}\times (\nabla\times{\mathbf
n})]^{2}-K_{24}\nabla\cdot[\mathbf n\cdot\nabla\cdot\mathbf
n+\mathbf n \times (\nabla \times \mathbf n )]^2 ].
\end{equation}
The magnetic energy density has the form
$-\frac{\chi_a}{2}(\textbf{n}\cdot\textbf{H})^2$, (where $\chi_a$ is
the anisotropy of magnetic susceptibility, and $\textbf{H}$ is the
magnetic field).

The terms at $K_1$, $K_2$, and $K_3$ elastic constants in
$\ref{long}$ mean splay, twist, and bend deformations of a bulk
nematic respectively, $\textbf{n}$ is the coordinate dependent
nematic director. The term at $K_{24}$ relates to saddle-splay
deformation mode \cite{PP_pre}. In this continuum, the tactoid
boojums were revealed by Kaznacheev \cite{Kazna2002} and by van der
Schoot \cite{PP_pre} practically identically, independently of one
another. The final result of tactoid classification is the existence
of four regimes of form is possible, which depend on anchoring
between the local director and the tactoid surface, and also on the
total tactoid volume [11]. Only at the week-coupled limit
($\gamma=0$), Kaznacheev found an equilibrium shape of a tactoid
[2], \cite{Kazna2003}, \textit{a fortiori} at $F_m=0$ and without
the terms of $K_{24}$-s in (2). At the limit (Fig. 1), the free
energy (1) is the almost non-analytical function on
$f(\alpha,\gamma)$ \cite{Kazna2002, Kazna2003}:
\begin{widetext}
\begin{eqnarray}\nonumber
4\pi(\sin\alpha-\alpha
\cos\alpha)+\pi(3\sin\alpha-3\alpha\sin\alpha-\alpha^2\sin\alpha)+\\
\pi\sin^3\alpha
\int_{-\infty}^{\infty}\frac{\sin\theta}{(\cosh\eta_{kazn}+\cos\alpha)}d\eta_{kazn}+
\frac{\pi}{36}[\sin \alpha(20+\cos \alpha)-3\alpha\cos \alpha(7+2
\sin^2\alpha)]
\end{eqnarray}
\end{widetext}
here $\theta$ is the parameter with the too long dependence of
$\alpha$, $\gamma$, $\eta_{kazn}$ \cite{Kazna2003}, the last term of
(4) corresponds to the magnetic energy at $\gamma \rightarrow 1$.
For $\gamma$, see Fig.1.

Nematic surface defects of the tactoids \cite{Volovik78} are of the
homotopic group $\pi_2(R,\widetilde{R})=P \times Q$, the defects of
the $P$ group are living only at the surface ($P$ group is the
kernel of the homomorphism $\pi_1(\widetilde{R})\rightarrow\pi_1(R)$
and consists of integers \cite{Lavrentovich_dyn}), and $Q$'s defects
are arrived from the interior. (Here $R$ and $\widetilde{R}$ denote
the space of degenerate states in the volume and the non-vanishing
states on the surface, which are arrived from the interior,
respectively). The interior may be inhabited by hedgehogs. All of
these point defects keep within the exact homotopic sequence
\cite{Volovik78}:
\begin{equation}
\pi_2(\widetilde{R}) \longrightarrow \pi_2(R) \longrightarrow
\pi_1(\widetilde{R}) \longrightarrow \pi_1(R).
\end{equation}
Boojums are characterized by topological charges $m$ and $n$
\cite{Lavrentovich_dyn}, which depend on a configuration of a
nematic director's field. Annihilation of the boojums of the
adjacent tactoids does not mean an influence of the raising
hedgehog's (in topology, they are not arbitrary floating to the
tactoid surface). Kurik and Lavrentovich \cite{Kurik} have mentioned
about some strings, connecting opposite boojums via a hedgehog in
nematic droplets, however, non-triviality of $\pi_1$ group hampered
the revealing of the droplet junction without the disclination
concept. However, in our case we reasonably ignore lacking
disclinations (see the conclusions by Balachandran et al.
\cite{prl84}).

Interaction scales are the 'dipole length'  $L_{dip}$, and the
'correlation length' $L_{\xi}$~\cite{M}, which are characterized an
action of the group of the order parameter. We assume $L_{dip}$ is
in connection with a long-axis of a tactoid.

In the Cartesian coordinates ($x,y,z$), the director field has the
configuration
$\textbf{n}=n(0,0,\frac{1-\cosh\eta_{kazn}\cos\xi}{\cosh\eta_{kazn}-\cos\xi})$,
where $\eta_{kazn}, \xi$ are the bispherical coordinates \cite{Kazna2002}.

Quite evidently, that tactoid system is provided by a gauge field
\cite{Polyakov} (and a field with $SU(2)$ symmetry). Concerning an
universality class of the system, take the $V_2O_5$ droplet surface
as belonging to $SO(3)$ group of rotations of the two-dimensional
sphere (here 'tactoid') $S^2$ \cite{M}. $U(1)$ will a group of
rotations around a droplet axis, which is agree closely with the
magnetic phase group of $^3He-A$ \cite{Bais}. $U(1)$'s winding is
realized of non-trivial topology of tactoids.

The $SO(3)$ and $SU(2)$ groups are locally isomorphic (as their Lie
algebras) and are connected by the homomorphism, $SO(3)\sim
SU(2)/Z_2$, where our $Z_2$ is the boojum's boundary condition.

In our standpoint, at the bulk junction, the group $SO(3)^n\times
U(1)^{2n}$ broken down to $SO(3)^{n-1}\times U(1)^{2n-1}$, where $n$
is a number of tactoids.

A model of the sol should involve the monopole solutions, according
to the theorem \cite{M} about requirement of their existence
($\pi_2(G/H)\longrightarrow \pi_1(H)$).

On the other hand, inasmuch as $\pi_1(H)=\underbrace{Z\otimes
Z\otimes Z\otimes...\otimes Z}_{2n}$, the $V_2O_5-H_2O$ sols are of
the group $G$. The tactoid annihilation may be described either by
non-Abelian or Abelian theory in depending on the global field
SU(2). Besides, we have to expect appearance of a compensative
vector field \cite{BShir}.

Here, an each tactoid, in correspondence to two poles (boojums) on a
tactoid surface, may contain two vortons with their tails (the wide
and "over-Witten's" definition for vortons see in  \cite{Blaha,
radu, vorton}, this is a kind of monopoles with the definite pair of
topological charges, vortex and azimuthal windings). Just as
vortices, they appear, if the order parameter has extra degrees of
freedom besides of the overall phase \cite{Ivanov, Cheshire}. In the
tactoid free energy, the terms of twisted deformations~\cite{PP_cm}
may play a role in these excitations. By introducing a necessary
parametrization, the free energy equation, analogous to
\cite{Kazna2002}, was proposed in~\cite{PP_cm}, where the free
parameters permit to be the non-commutative relations in the droplet
symmetry. Let us note, that we use the factor-space $CP^1$ in
accordance to a chiral (gauge) field (\ref{long}) \cite{Polyakov}.

Though, due to the electromagnetic (no topological) reasons, the sol
tactoids can survive coalescence owing to the Coulomb attraction in
water. But from topology \cite{Kurik}, we do not yet know about
appearance of a physical field from the configuration of defects. We
have to note, that because of in-homogeneity of a system, we have a
wide class of string models for a prototype.

\section{Annihilation of topological defects}
So, a junction of droplets means, that the surface point defect
(boojum) configuration may be unstable ($\gamma\neq0$). We discuss
the Abelian and non-Abelian string configurations \cite{Bais,
BaisAA, Morris, Yung_004, Oshikawa2}, which support the sols of
tactoid nematics. Their combinations and interactions are expected
to define of the junction of tactoids.

\subsection{Abelian space}
The Abelian character of pair boojums and monopoles, and also their
integer charge were proven \cite{Volovik2}. Boojums of charge
$N=\pm1$ live at $L_{\xi} \ll L \ll L_{dip}$~\cite{M, Blaha}. But
from the surface field phenomenology \cite{Kazna2002} of a solitary
tactoid, one can not define a flux number $k$~\cite{M}, concerning
an each boojum, only what $k=1$ is preferable for their pairing
configuration, and $k=2$ for a unit singularity. In this scenario,
annihilation of charge-opposite (topological) 'particles' is
possible.

Abelian monopoles may be associated with locations of boojums, but,
due to the topological properties of our G, we ignore them. Let us
consider only vortons of the Abelian gauge. They are unstable
\cite{Morris}, and appear together with the neutral strings. The
open question is which velocity will greater: of the tactoid
coalescence or the vorton decay.

In the U(1) gauge, the loop-radius dependent criterion of the vorton
stability was found and analyzed  numerically in the case of the
potential expressed in the elliptic ansatz [27], as well as in the
well-known Witten's $U(1)\times U(1)$ case (see review [21), that is
an analogous phase transition from $U(1)\times U(1)$ to $U(1)$ for
two neighbouring randomly oriented tactoids, in absent of magnetic
field.

\subsection{Non-Abelian space}
Usual Lagrangians of non-Abelian theories are often linearized into
the Bogomolny-Prasad-Sommerfeld (BPS) equations \cite{Bogomolny}. A
number of applications corresponding to similar strings were
considered, for example, in \cite{prl84, radu, vorton, Cheshire,
BaisAA, Yung_004, Oshikawa2, Janih, LoPr}.

In the phase diagram \cite{PP_cm}, the regions of twist states were
indicated. If the tactoid junction carry out there, for spherical
and prolate droplets, one may make an analogy between the
non-Abelian vortons and "rotation"  of the nematic order parameter,
in spite of the ansatz ($\alpha(\eta)=\alpha_0\sin\eta$
\cite{PP_cm}) condition, labeled one of the topological invariants.

Let us formulate the string model with the boson Lagrangian density
(due to \cite{radu, vorton})
\begin{equation}
{\mathfrak{L}}=-\frac{1}{4}F_{\mu\nu}F^{\mu\nu}-\frac{1}{4}\textbf{G}_{\mu\nu}\textbf{G}^{\mu\nu}\\
-D_{\mu}\vec{\phi}^\dag \cdot
D^{\mu}\vec{\phi}-V(\vec{\phi}).
\end{equation}
Here
\begin{equation}
F^{\mu\nu}=\partial_\mu A_\nu-\partial_\nu A_\mu
\end{equation}
are the Abelian field strengths. The global curvature is
\begin{equation}
G_{\mu\nu}=\partial_\mu \mathbf H_ \nu -\partial_\nu \mathbf H_\mu+g \mathbf H_\mu\times \mathbf H_\nu.
\end{equation}
The gauge covariant derivatives of vacuums are:
\begin{equation}
D_{\mu}(\vec{\phi})=\partial_\mu
\vec{\phi}-ieA_{\mu}\vec{\phi}+g\mathbf H_\mu \times \vec\phi,
\end{equation}
In the formulas (5) - (8), $\mu$ and $\nu$ are indices of the gauge
field $A$ and of the metrics $g$. $H_\mu$ and $\phi$ are the
three-dimensional vectors in the $SU(2)$ Lie algebra. The field
potential $V(\mathbf \phi)$ is expressing from (2). Due to
\cite{Kazna2003}
\begin{equation}
x=a \frac{\sin\xi\cos\varphi}{\cosh \eta -\cos \xi},
y=a \frac{\sin \xi \sin \varphi}{\cosh \eta -\cos \xi},
z=a \frac{\sinh \eta}{\cosh \eta-\cos \xi},
\end{equation}
 the bulk elastic energy \cite{Kazna2003} of a tactoid equals to
\begin{equation}
\frac{a\gamma}{2}\int_0^{2\pi}d\varphi\int_{-\infty}^{\infty}d\eta_{kazn}\int_{\pi
-\alpha}^{\pi}\frac{4K_1\sinh^2\eta_{kazn}\sin\xi+K_3\sin^3\xi}{(\cosh\eta_{kazn}-\cos\xi)^3}Dd\xi.
\end{equation}
The corresponding vector potential is
\begin{equation}
V(\vec{\phi})=\frac{1}{2}\lambda+(\vec{\phi}^{+}\cdot
\vec{\phi}-\frac{1}{2}\zeta^2)^2+\frac{1}{2}k|\vec{\phi} \cdot
\vec{\phi}|^2.
\end{equation}
At the parameter $k>0$, the vacuum is characterize by $\vec{\phi}
\cdot \vec{\phi}$,
$\vec{\phi}^\dag\cdot\vec{\phi}=\frac{1}{2}\zeta^2$.
\begin{equation}
 \mathbf \Phi_0 = \frac{\zeta}{2}
\left (
 \begin{array}{ccc}
-\frac {\sin \xi \sinh \eta_{kazn} \cos \varphi_{kazn}}{\cosh \eta_{kazn} - \cos \xi} \\
-\frac {\sin\xi\sinh\eta_{kazn}\sin\varphi_{kazn}}{\cosh\eta_{kazn}-\cos\xi}  \\
\frac {1-\cosh_{kazn}\cos\xi}{\cosh\eta_{kazn}-\cos\xi}
 \end{array} \right ).
\end{equation}
The generators of $SU(2)$ are denoted as $T_{i=1,2,3}$. $T_0$ is the
generator of $U(1)$.
$-iT_1(\vec{\phi}_j)=-\epsilon_{ijk}\vec{\phi}_j$,
$-iT_0(\vec{\phi})_j=-\vec{\phi}_j$. $Q=T_2+T_0$ is the annihilation
condition. The string generator ($T_S=T_3$) does not commutate with
the charge generator: $[T_S,Q]=[T_3,T_2]=-iT_1$ \cite{vorton}. Here
$\vec{\phi}(\alpha)=e^{-i\alpha T_3}\vec{\phi}$ are also meaning the
generators. Between the vacuums, the angular dependence is
established $Q(\theta)=e^{-i\theta T_s}Qe^{i\theta T_s}$ [29].
Tactoid vortices revolve $SU(2)$. $R$, $\alpha$ are introduced to
describe the tactoid geometry (Fig.1).

Further, we need to solve the next equations of motion:
\begin{equation}
\frac{1}{\sqrt{g}}\partial_\mu\sqrt{g}\mathbf F ^{\mu\nu}=j^{\alpha}=je[\vec{\phi}^\dag\times D^\nu\vec{\phi}-\vec{\phi}\cdot(D^\nu \vec{\phi})^\dagger ],
\end{equation}
\begin{equation}
\frac{1}{\sqrt{g}}\partial_\mu\sqrt{g}\mathbf G ^{\mu\nu}=\mathbf J^{\alpha}=g[\vec{\phi}^\dag\times D^\nu\vec{\phi}+\vec{\phi}\times(D^\nu \vec{\phi})^\dagger ],
\end{equation}
\begin{equation}
\frac{1}{\sqrt{g}}D_\mu\sqrt{g}\vec{\phi}=\frac{\delta V}{\delta \vec{\phi}^\dag}.
\end{equation}
To confirm the existence of vortons, labeled by vacuum, and estimate
the energy $T_2$, the first-order Bogomol'ny's equations are usually
applied. the first-order Bogomolny's equations are usually applied.
For example, in the sigma-model limit of the Lagrangian of the type
(5), the non-Abelian votrons with the (1, 1)-, (1, 2)- and other
pairs of winding numbers in $SU(2)$ were numerically revealed by
Radu and Volkov [21] just lately; to be solvable, their model has
included four free parameters in the potential (Fig. 2).

There was numerically proven with help of Gauss-Tschebuchev
algorithm, that in the $U(1)$ gauge, the stable vortons may appear
[27], whereas in $SO(3)$ it is not so [26]. The stability criterion
includes the radius $R$ of the vortex loop, which may be compared
with the Kaznacheev-van der Schoot theoretical analysis [2], [7],
and with the lattice Monte Carlo simulations, performed by Bates
[11].

\section{Dynamics and estimations for time-aging of the tactoid sols}
Along with these assertions on the configurations supplied with
Non-Abelian gauge fields, the approximate methods of analysis exist
for quite attainable numerical simulations of vorton states. One of
there is so called Abelian projection [33]. So, following the
Maximal Abelian (MaA) projection approach, we fix $SU(2)$ gauge and
leave the winding group $U(1)$ unfixed. In applied numerical tasks,
Abelian approximations of (11) are yet acceptable.

For example, whether is an analogous Abelian projection of the
$V_2O_5-H_2O$ tactoid configuration realized in the $2D$
ferromagnetic systems and thin films [34], if there are defined the
same topological invariants? This simplification is useful to
estimate the case of annihilating particles with whole unit opposite
charges [3]. One may express the vorton dynamics by the
Landau-Lifshitz equation (LLE), including dissipation (labeled by
the constant ad [34]), and write:

\begin{figure}
\includegraphics*[width=70mm]{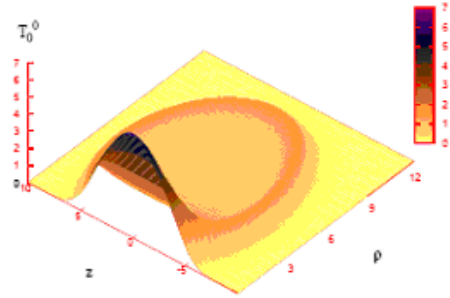}
\caption{\small The energy density of n=m=1 vortons [21], plotted by
Radu and Volkov numerically at four free parameters, where z and ?
are the polar coordinates.}
\end{figure}
\begin{figure}
\includegraphics*[width=70mm]{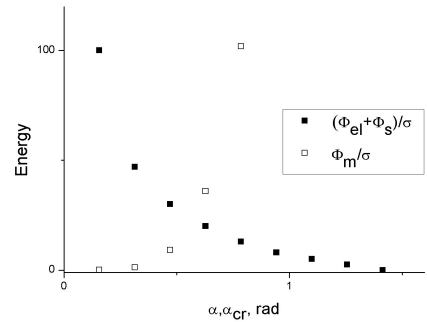}
\caption{\small Competition between the magnetic $\Phi_{m}$ and the
elastic and the surface energies $\Phi_{el} + \Phi_{s}$ of a
tactoid, divided by $\sigma=10^{-3}$ erg/cm, errors are not
indicated; as this is a qualitative view of (3) and experiments [2],
[12]; the data at $\gamma\rightarrow 1$, from which $\alpha\approx
32^o$.}
\end{figure}

\begin{equation}
\frac{\partial\textbf{m}}{\partial
t}=\textbf{m}\times\textbf{f}-\alpha_d\textbf{m}\times(\textbf{m}\times\textbf{f}),
\textbf{f}\equiv\triangle\textbf{m}-Qm_z\hat{\textbf{e}}_z,
\end{equation}
where $\textbf{m}$ is the magnetization vector, $Q$ is the free
parameter, $\hat{\textbf{e}}_z$ is the unit vector in the $z$
magnetization direction. According to the definition [3], the
topological invariant $N$ connected with the topological density $n$
is $N=\frac{1}{4\pi}\int_Vn\epsilon_{\mu\nu}\textbf{r}^3$,
$\epsilon_{\mu\nu}$ is the asymmetric tensor with $(\mu,
\nu)=(1,2)$, $V$ is a tactoid volume, and the vector $\textbf{r}$
denotes its space.

The magnetic field stretches large tactoids (a increases), whereas
to annihilate, the tactoid shape should become more oblate [12],
Fig. 3. Therefore, the equilibrium angle a, corresponding to the
large tactoid shape, exists also for coalescence in the magnetic
field. To define a is not difficult from the next simple algebra
with (1) and (2), by using the definitions [3, 10-11]. From (1) -
(3) and (12), dynamics characteristics of a solitary tactoid may be
expressed as:
\begin{equation}
\frac{1}{(\chi H)^4}(\frac{\partial\textbf{m}}{\partial t})^2\sim
\alpha_d^2(\frac{1-\cosh\eta_{kazn}\cos\xi}{\cosh\eta_{kazn}-\cos\xi})^2.
\end{equation}
E. g. stretch of a tactoid in $z$-axis direction increases its
magnetic energy, and the magnetic field is precipitating for
annihilation of droplets, as a free volume decreases.

On the experiment [14], the next parameters are measured:
$C_i=\frac{K_i}{\sigma}$, $i=1,3$, $K_i$ are modulii of (2), and
$\sigma$ is the surface tension. $C_3$-s order is hundreds
micrometers. For $C_1$-s, these are about unit. Both of they are
drop-down with time, but according to (1)-(2), have not affect on
the magnetic term.

\section{Conclusions}
We composed the topological classification of sols $V_2O_5-H_2O$,
owing to which, the qualitative practical predictions for
thermodynamic states of these sols may be performed. The
cosmological theory of superconductive strings supposes that the
nematic tactoids in $V_2O_5-H_2O$ annihilate in accordance with
non-Abelian statistics.

This process, carried out in magnetic field, increases a time-aging
of the sols, but does not yield to direct exact estimations, since
its nature is principally Non-Abelian. One may connect an actual
electromagnetic interaction in the $V_2O_5-H_2O$ solution via $pH$
value and discuss questions on the tactoid junction in frames of
chemistry, which we have wittingly ignored in favor of the important
topological role. The process of tactoid junction in magnetic field
leads to rise of the additional electromagnetic field changing $pH$
of water around tactoids and, for one's part, time-aging [2, 13].
These observations may be important for ecology, as long as vanadium
pentoxide is contained in impurities of coal soles, which are the
components of wastes of thermoelectric power stations and are
included in the impurity parameters at the background control for
radiation.
$$
$$

\end{document}